\documentclass[12pt,fleqn]{article}
\usepackage{amsfonts,cite}
\usepackage[dvips]{graphics}

\def\onecol{\onecolumn \mathindent 2em}

\def\noi{\noindent}

\makeatletter
\renewcommand{\section}{\@startsection{section}{1}{0pt}%
        {-3.5ex plus -1ex minus -.2ex}{2.3ex plus .2ex}%
        {\large\bf\protect\raggedright}}

\renewcommand{\subsection}{\@startsection{subsection}{2}{0pt}%
        {-3ex plus -1ex minus -.2ex}{1.4ex plus .2ex}%
        {\normalsize\bf\protect\raggedright}}

\renewcommand{\thesubsubsection}%
        {\arabic{section}.\arabic{subsection}.\arabic{subsubsection}.}
\renewcommand{\@oddhead}{\raisebox{0pt}[\headheight][0pt]{%
   \vbox{\hbox to\textwidth{\rightmark \hfil \rm \thepage \strut}\hrule}}}
\renewcommand{\@evenhead}{\raisebox{0pt}[\headheight][0pt]{%
   \vbox{\hbox to\textwidth{\thepage \hfil \leftmark \strut}\hrule}}}

\makeatother

\newcommand{\Title}[1]{\noi {{\Large\bf #1}} \\}

\def\Aunames#1{\noi{\large\bf #1}}
\def\auth#1{${}^{#1}$}
\def\Addresses#1{\medskip\noi \protect
    \begin{description}\itemsep -3pt
        {\it #1} \end{description}}
\def\addr#1#2{\item[${}^{#1}$]{\it #2}}
\newcommand{\Abstract}[1]{\vskip 2mm \begin{center}
        \parbox{16.4cm}{\small\noi #1} \end{center}\medskip}

\newcommand{\email}[2]{\footnotetext[#1]{e-mail: #2}
        \addtocounter{footnote}{1}}

\def\nqq{\hspace*{-2em}}
\def\nhq{\hspace*{-0.5em}}

\def\cm{\hspace*{1cm}}
\def\inch{\hspace*{1in}}

\def\Jl#1#2{#1 {\bf #2},\ }

\def\ApJ#1 {\Jl{Astroph. J.}{#1}}
\def\CQG#1 {\Jl{Class. Quantum Grav.}{#1}}
\def\DAN#1 {\Jl{Dokl. AN SSSR}{#1}}
\def\GC#1 {\Jl{Grav. Cosmol.}{#1}}
\def\GRG#1 {\Jl{Gen. Rel. Grav.}{#1}}
\def\JETF#1 {\Jl{Zh. Eksp. Teor. Fiz.}{#1}}
\def\JETP#1 {\Jl{Sov. Phys. JETP}{#1}}
\def\JHEP#1 {\Jl{JHEP}{#1}}
\def\JMP#1 {\Jl{J. Math. Phys.}{#1}}
\def\NPB#1 {\Jl{Nucl. Phys.}{B\ #1}}
\def\NP#1 {\Jl{Nucl. Phys.}{#1}}
\def\PLA#1 {\Jl{Phys. Lett.}{#1A}}
\def\PLB#1 {\Jl{Phys. Lett.}{#1B}}
\def\PRD#1 {\Jl{Phys. Rev.}{D\ #1}}
\def\PRL#1 {\Jl{Phys. Rev. Lett.}{#1}}

\def\al{&\nhq}
\def\lal{&&\nqq {}}
\def\eq{Eq.\,}
\def\eqs{Eqs.\,}
\def\beq{\begin{equation}}
\def\eeq{\end{equation}}
\def\bear{\begin{eqnarray}}
\def\bearr{\begin{eqnarray} \lal}
\def\ear{\end{eqnarray}}
\def\earn{\nonumber \end{eqnarray}}
\def\nn{\nonumber\\ {}}

\def\nnn{\nonumber\\ \lal }

\def\yy{\\[5pt] {}}
\def\yyy{\\[5pt] \lal }
\def\eql{\al =\al}

\def\tst{\textstyle}

\def\fract#1#2{{\tst\frac{#1}{#2}}}

\def\half{{\fract{1}{2}}}

\def\e{{\,\rm e}}
\def\d{\partial}

\def\im{\mathop{\rm Im}\nolimits}

\def\sign{\mathop{\rm sign}\nolimits}

\def\const{{\rm const}}
\def\eps{\varepsilon}

\def\then{\ \Rightarrow\ }

\def\DAL{\mathop{\raisebox{3.5pt}{\large\fbox{}}}\nolimits\,}
\newcommand{\vars}[1]{\left\{\begin{array}{ll}#1\end{array}\right.}

\def\ep{\epsilon}
\def\eps{\varepsilon}

\def\ok{{\overline k}{}}
\def\ot{{\overline t}{}}
\def\ou{{\overline u}{}}

\def\R{{\mathbb R}}
\def\df{\delta\phi}
\def\da{\delta\alpha}
\def\db{\delta\beta}
\def\dg{\delta\gamma}

\def\sph{spherically symmetric}
\def\ssph{static, spherically symmetric}
\def\asflat{asymptotically flat}
\def\wh{wormhole}
\def\whs{wormholes}
\def\bh{black hole}
\def\bhs{black holes}
\def\afi{anti-Fisher}

\def\Schr{Schr\"odinger}

\def\GR{general relativity}

\def\mn{_{\mu\nu}}

\def\mN{_\mu^\nu}

\def\Veff{V_{\rm eff}}
\def\Wreg{W_{\rm reg}}
\def\uth{u_{\rm th}}

\begin{document}
\thispagestyle{empty}
\onecol

\Title {On the stability of scalar-vacuum space-times}

\Aunames{\normalsize K.A. Bronnikov,\auth{a,1} J.C. Fabris,\auth{b,2}
     and A. Zhidenko\auth{c,3}}

\Addresses{\small
\addr a {Center for Gravitation and Fundamental Metrology, VNIIMS,
     Ozyornaya 46, Moscow 119361, Russia;\\
     Institute of Gravitation and Cosmology,
     PFUR, ul. Miklukho-Maklaya 6, Moscow 117198, Russia}
\addr b {Departamento de F\'{\i}sica, Universidade Federal do
     Esp\'{\i}rito Santo,\\ avenida Fernando Ferrari 514,
     29075-910 Vit\'oria, ES, Brazil}
\addr c {Centro de Matem\'atica, Computa\c{c}\~ao e Cogni\c{c}\~ao,
     Universidade Federal do ABC,\\
         Rua Santa Ad\'elia, 166, 09210-170, Santo Andr\'e, SP, Brazil}
    }

\Abstract
  {We study the stability of \ssph\ solutions to the Einstein equations
   with a scalar field as the source. We describe a general methodology of
   studying small radial perturbations of scalar-vacuum configurations with
   arbitrary potentials $V(\phi)$, and in particular space-times with
   throats (including wormholes), which are possible if the scalar is
   phantom. At such a throat, the effective potential for perturbations
   $\Veff$ has a positive pole (a potential wall) that prevents
   a complete perturbation analysis. We show that, generically, (i) $\Veff$
   has precisely the form required for regularization by the known
   S-deformation method, and (ii) a solution with the regularized potential
   leads to regular scalar field and metric perturbations of the initial
   configuration. The well-known conformal mappings make these results
   also applicable to scalar-tensor and $f(R)$ theories of gravity. As a
   particular example, we prove the instability of all static solutions with
   both normal and phantom scalars and $V(\phi) \equiv 0$ under spherical
   perturbations. We thus confirm the previous results on the unstable
   nature of anti-Fisher \whs\ and Fisher's singular solution and prove the
   instability of other branches of these solutions including the
   anti-Fisher ``cold \bhs''.  }

\email 1 {kb20@yandex.ru}
\email 2 {fabris@pq.cnpq.br}
\email 3 {olexandr.zhydenko@ufabc.edu.br}

\section {Introduction}

  Phantom field configurations have become a subject of particular interest
  since the discovery of the accelerated expansion of our Universe
  and its explanation in the framework of general relativity by the
  existence of dark energy (DE), a source of gravity of unknown nature with
  the pressure to density ratio $w < -1/3$. Numerous observations lead to
  estimates of $w$ around $-1$, which corresponds to a cosmological
  constant, but values smaller than $-1$ are still admissible and, moreover,
  preferable for describing an increasing acceleration. The most recent
  estimates read $w = -1.10 \pm 0.14$ ($1\,\sigma$) \cite{komatsu}
  (according to the 7-year WMAP data) and $w = -1.069^{+0.091}_{-0.092}$
  \cite{sullivan} (mainly from data on type Ia supernovae from the
  SNLS3 sample).

  The possible existence of phantom DE is certainly only one of potentially
  viable explanations of the observations (numerous DE models are described,
  e.g., in the reviews \cite{DE-rev1,DE-rev2,DE-rev3,DE-rev4,DE-rev5}),
  but if we take it as a working hypothesis, it is natural to expect that
  there are manifestations of DE in local objects and phenomena. The
  simplest of them can be described by \ssph\ solutions to Einstein-scalar
  equations where the scalar field has an unusual sign of kinetic energy (a
  phantom scalar, by definition). In the case of a massless scalar it is the
  so-called anti-Fisher solution, a phantom analogue of Fisher' solution
  \cite{fisher} for an ordinary minimally coupled massless scalar field. The
  anti-Fisher solution consists of three branches, one of them, termed
  Branch C in this paper, represents \whs\ \cite{h_ellis, br73}, the others,
  A and B, also have properties of interest, and some of them describe
  so-called cold black holes \cite{we99, we-afish}, which possess horizons
  of infinite area and zero Hawking temperature; in all such solutions there
  are throats (minima of the spherical radius), but, unlike \whs, here
  beyond the throat one does not find an \asflat\ region.

  A number of solutions for phantom scalar fields with nonzero
  self-interaction potentials are also known, among them are wormholes with
  flat and AdS asymptotic behaviours and a specific class of regular \bhs\
  called black universes \cite{pha1,pha4}. These are configurations where a
  possible explorer, after crossing the event horizon, gets into an
  expanding universe instead of a singularity, and moreover, this expanding
  universe, being initially highly anisotropic, eventually isotropizes and
  approaches a de Sitter mode of expansion at late times. The stability of
  all such models under small perturbations is an important test of their
  possible viability \cite{Konoplya:2011qq}.

  The purpose of this paper is to present a methodology of studying small
  radial perturbations of scalar-vacuum configurations with any potential
  $V(\phi)$ (see (\ref{L})), and in particular space-times with throats. The
  difficulty with the latter consists in the fact that the effective potentials
  $\Veff(x)$ for perturbations (not to be confused with the self-interaction
  potential $V(\phi)$) always, in any gauge, possess a singularity at the
  throat, which prevents a complete perturbation analysis. It is for this
  reason that some previous stability studies of anti-Fisher
  solutions \cite{we99, picon02} did not find any unstable mode whereas a
  numerical perturbation analysis of Shinkai and Hayward \cite{shin02}
  revealed an instability of the simplest representative of this family of
  solutions, the Ellis massless \wh\ \cite{h_ellis, br73}.

  Gonzalez et al. \cite{gon08}, analyzing the stability of anti-Fisher \whs,
  made a proper substitution in the perturbation equation (a special case of
  the so-called S-deformation \cite{kod03, kod11}) and regularized the
  effective potential $\Veff$. As a result, they found an exponentially
  growing mode with nonzero perturbation of the throat radius, thus showing
  that the anti-Fisher \whs\ are unstable. We show here that a similar
  methodology can be applied to more general self-gravitating scalar field
  configurations including those with arbitrary self-interaction potentials
  $V(\phi)$. To this end, we prove that, generically, (i) the effective
  potential $\Veff$ has precisely the form required for regularization by
  S-deformation and (ii) any solution of the transformed wave equation with
  a regularized potential leads to a regular perturbation of the background
  static configuration.

  As a particular example, we study the stability of all \afi\ solutions
  under \sph\ perturbations. We prove the instability of Branch A and B
  solutions and confirm the conclusions of \cite{gon08} for Branch C
  solutions (\whs).

\section {Perturbation equations}

\subsection {Preliminaries}

  Consider a self-gravitating, minimally coupled scalar field with an
  arbitrary self-interaction potential in \GR. The Lagrangian is
  (up to a constant factor)
\bear                                                       \label{L}
     L = \sqrt{-g}\biggr(R + \ep
          g^{\alpha\beta}\phi_{;\alpha}\phi_{;\beta} - 2V(\phi)\biggr),
\ear
  where $\ep =1$ for a normal scalar field with positive kinetic energy
  and $\ep=-1$ for a phantom scalar field. Other notations are usual,
  the gravitational constant is absorbed in the definitions of $\phi$
  and $V(\phi)$. The field equations are
\bearr                                                      \label{SE}
      \ep \DAL \phi + V_\phi =0,
\\ \lal                                                     \label{EE}
      R\mN = -\ep \phi_{,\mu}\phi^{,\nu} + \delta\mN V(\phi),
\ear
  where $V_\phi \equiv dV/d\phi$.

  The general \sph\ metric may be written in the form
\beq                                                         \label{ds}
    ds^2 = g\mn dx^{\mu}dx^{\nu}
       = \e^{2\gamma} dt^2 - \e^{2\alpha}du^2 - \e^{2\beta}d\Omega^2,
\eeq
  where $\alpha$, $\beta$, $\gamma$ are functions of the radial coordinate
  $u$ and the time coordinate $t$ and $d\Omega^2= d\theta^2 + \sin^2\theta\;
  d\varphi^2$. We will also use the notation $r = \e^\beta$ for the areal
  radius, such that $4\pi r^2$ is the area of coordinate spheres
  $u=\const,\ t=\const$. There remains a coordinate freedom in the $(u,t)$
  subspace, which in general corresponds to choosing a reference frame
  preserving the spherical symmetry. In the static case, there is a
  reference frame such that there is no $t$-dependence, and then the
  coordinate freedom concerns the choice of the $u$ coordinate.

  Curvature singularities for \ssph\ space-times are entirely determined
  using the Kretschamnn scalar, which, for the metric (\ref{ds}), may be
  written as a sum of squares,
\beq                                                        \label{k}
   K = 4K_1^2 + 8K_2^2 + 8K_3^2 + 4K_4^2,
\eeq
  with
\bear                                                       \label{k_i}
      K_1 \eql {R^{01}}_{01} = - \e^{-\alpha - \gamma}
                \Bigr(\gamma'\e^{\gamma - \alpha}\Bigl)',
\nn
      K_2 \eql {R^{02}}_{02} = {R^{03}}_{03} = - \e^{-2\alpha}\beta'\gamma',
\nn
      K_3 \eql {R^{12}}_{12} = {R^{13}}_{13} = - \e^{-\alpha - \beta}
                \Bigr(\beta' \e^{\beta-\alpha}\Bigl)',
\nn
      K_4 \eql {R^{23}}_{23} = \e^{-2\beta} - \e^{-2\alpha}{\beta'}^2,
\ear
  where the prime denotes $d/du$. The structure of \eq(\ref{k}) indicates
  that an infinite value of any $K_i$ at some value of $u$ implies the
  presence of a singularity at this $u$. All these expressions for $K_i$ are
  invariant with respect to the choice of the $u$ coordinate.

  We will assume that a certain \ssph\ solution to \eqs (\ref{SE}) and
  (\ref{EE}) is known and study its stability under small \sph\
  perturbations. We thus consider, instead of $\phi(u)$, a perturbed unknown
  function
\[
    \phi(u,t)= \phi(u)+ \df(u,t)
\]
  and similarly for the metric functions $\alpha,\ \beta,\ \gamma$, where
  $\phi (u)$, etc., are taken from the static solutions.

  Preserving only linear terms with respect to time derivatives, we can
  write all the nonzero component of the Ricci tensor and the time-time
  component of the Einstein tensor as
\bear
     R^0_0 \eql                                         \label{R00}
     \e^{-2\gamma}(\ddot\alpha + 2\ddot\beta)
           -\e^{-2\alpha}[\gamma'' +\gamma'(\gamma'-\alpha'+2\beta')],
\\
     R^1_1 \eql
     \e^{-2\gamma}\ddot\alpha                           \label{R11}
     - \e^{-2\alpha}[\gamma''+2\beta'' +\gamma'{}^2+2\beta'{}^2
            -\alpha'(\gamma'+2\beta')],
\\
     R^2_2 \eql \e^{-2\beta}                                    \label{R22}
          +\e^{-2\gamma}\ddot\beta
              -\e^{-2\alpha}[\beta''+\beta'(\gamma'-\alpha'+2\beta')],
\\
     R_{01}\eql
        2[\dot\beta' + \dot{\beta}\beta'                        \label{R01}
                 -\dot{\alpha}\beta'-\dot{\beta}\gamma'],
\\                              \label{G00}
     G^0_0 \eql \e^{-2\alpha}[2\beta''+\beta'(3\beta'-2\alpha')]
                        -\e^{-2\beta},
\ear
  where dots and primes denote $\d/\d t$ and $\d/\d u$, respectively.

\subsection {General form of the field equations}

  The zero-order (i.e., static) scalar, ${0\choose 0}$, ${1\choose 1}$,
  ${2\choose 2}$ components of \eqs (\ref{EE}) and the Einstein equation
  $G^0_0 = \ldots$ are
\bear                                                      \label{e-phi0}
     \phi'' + \phi'(\gamma'+2\beta'-\alpha') \eql \ep \e^{2\alpha}V_\phi,
\yy                                                        \label{00-0}
     \gamma'' + \gamma'(\gamma'+2\beta'-\alpha') \eql -V\e^{2\alpha},
\yy                                                         \label{11-0}
     \gamma'' + 2\beta'' + \gamma'{}^2 + 2\beta'{}^2
            -\alpha'(\gamma'+2\beta') \eql  -\ep \phi'^2 - V\e^{2\alpha};
\yy                                                        \label{22-0}
     -\e^{2\alpha-2\beta}
       + \beta'' + \beta'(\gamma'+2\beta'-\alpha') \eql -V\e^{2\alpha},
\yy                                            \label{G00-0}
     -\e^{2\alpha-2\beta} + 2\beta''+\beta'(3\beta'-2\alpha')
                \eql -\half \ep \phi'{}^2 - V \e^{2\alpha}.
\ear
  The first-order perturbed equations (scalar, $R_{01}=\ldots$,
  $R^2_2 = \ldots$, and $G^0_0 = \ldots$) read
\bearr                                                     \label{e-phi1}
     \e^{2\alpha-2\gamma} \delta\ddot\phi -\df''
        -\df' (\gamma'+2\beta'-\alpha')
            -\phi'(\dg' + 2\db' -\da')
            + \ep \delta(\e^{2\alpha}V_\phi) =0,
\yyy                                                       \label{01-1}
    \delta\dot\beta' + \beta'\delta\dot\beta
        - \beta' \delta\dot\alpha - \gamma' \delta\dot{\beta}
                = - \half \ep \phi'\delta\dot\phi,
\yyy                                                       \label{22-1}
    \delta(\e^{2\alpha-2\beta})
            + \e^{2\alpha-2\gamma} \delta\ddot\beta
                    -\db'' -\db'(\gamma'+2\beta'-\alpha')
\nnn \inch\inch
        -\beta'(\dg' + 2\db' -\da')  = \delta(\e^{2\alpha} V),
\yyy                                           \label{G00-1}
    -\delta(\e^{2\alpha-2\beta})
            + \db'' + 6\beta' \db' - 2\beta' \da' -2\alpha' \db'
                    = -\ep \phi' \df - \delta(V\e^{2\alpha}).
\ear
  \eq (\ref{01-1}) may be integrated in $t$; since we are interested in
  time-dependent perturbations, we omit the appearing arbitrary
  function of $u$ describing static perturbations and obtain
\beq                                                       \label{01-1i}
    \db' + \db(\beta'-\gamma') - \beta'\da = -\half\ep\phi' \df.
\eeq

  Let us note that we have two independent forms of arbitrariness: one is the
  freedom of choosing a {\it radial coordinate\/} $u$, the other is a {\it
  perturbation gauge\/}, or, in other words, a reference frame in the
  perturbed space-time, which can be expressed in imposing a certain
  relation for $\da,\ \db$, etc. In what follows we will employ both kinds
  of freedom.  All the above equations have been written in the most
  universal form, without coordinate or gauge fixing.

\subsection {Gauge $\db \equiv 0$}

  This is technically the simplest gauge, in particular, it is convenient
  for considering usual \bh\ perturbations, but causes certain difficulties
  when applied to \whs\ and other configurations with throats. The reason
  is that the assumption $\delta\beta = 0$ leaves invariable the
  throat radius, while perturbation must in general admit its time
  dependence \cite{we99}. This problem will be discussed below.

  With $\db = 0$, \eq (\ref{01-1i}) expresses $\da$ in terms of $\df$:
\beq                                  \label{da-df}
    2 \beta' \da = \ep \phi' \df.
\eeq
  \eq (\ref{22-1}) expresses $\dg' - \da'$ in terms of $\da$ and $\df$:
\beq                                  \label{dg-df}
    \beta'(\dg'-\da') = 2 \e^{2\alpha-2\beta}\da
                        - \delta(\e^{2\alpha} V).
\eeq
  Substituting all this into (\ref{e-phi1}), we obtain the wave equation
\bearr                                                        \label{eq-df}
      \e^{2\alpha-2\gamma} \delta\ddot\phi
            -\df'' - \df' (\gamma'+2\beta'-\alpha') + U \df =0,
\nnn  \cm
      U \equiv  \e^{2\alpha}\biggl\{
     \ep (V - \e^{-2\beta})\frac{\phi'^2}{\beta'^2}
          + \frac{2\phi'}{\beta'} V_\phi + \ep V_{\phi\phi}\biggr\}.
\ear

  Before proceeding with a study of the wave equation, let us make sure
  that all the remaining Einstein equations hold as a consequence of
  (\ref{da-df}), (\ref{e-phi1}) and (\ref{22-1}) and do not lead to any new
  restrictions.  Consider the component (\ref{G00-1}) (the constraint
  equation). It now takes the form
\beq                                             \label{G00-1b}
     2\da\,\e^{2\alpha-2\beta} + 2\beta' \da'
            = \ep \phi' \df + \delta(V\e^{2\alpha}).
\eeq
  This equation holds automatically owing to the zero-order equations and
  (\ref{da-df}). Indeed, a substitution of $\da$ from (\ref{da-df}) brings
  (\ref{G00-1b}) to the form
\beq                                            \label{1b'}
      \frac {\df}{\beta'}\biggl[\beta'\phi''-\phi'\beta''
    + \e^{2\alpha-2\beta}\phi' - \e^{2\alpha}V \phi'
            - \ep \e^{2\alpha} V_\phi \beta' \biggr] = 0,
\eeq
  Now, substituting $\phi''$ and $\beta''$ from (\ref{e-phi0}) and
  (\ref{22-0}), respectively, we see that all terms cancel, i.e., the
  equation does hold. Furthermore, the Einstein equation $G^1_1 = \ldots$
  holds as a consequence of (\ref{22-1}) and (\ref{G00-1}); lastly, the
  equation $G^2_2 = \ldots$ holds due to the Bianchi identity $\nabla_\nu
  G^\nu_1=0$ and the corresponding property of the stress-energy tensor of
  the scalar field.

  So we can return to \eq (\ref{eq-df}).
  Passing on to the ``tortoise'' coordinate $x$ introduced according to
\beq
       du/dx = \e^{\gamma-\alpha}                           \label{to_x}
\eeq
  and changing the unknown function $\df \mapsto \psi$ according to
\beq                                                  \label{to_psi}
       \df = \psi(x,t) \e^{-\beta}, \ \ \
            \Leftrightarrow \ \ \ \psi(x,t) = r \delta\phi,
\eeq
  we reduce the wave equation to its canonical form, also called
  the master equation for radial perturbations:
\beq                                                        \label{wave}
       \ddot \psi - \psi_{xx} + \Veff (x)\psi =0,
\eeq
  (the index $x$ denotes $d/dx$), with the effective potential
\bearr                                                     \label{Veff}
     \Veff (x) = \e^{2\gamma-2\alpha}
            [U + \beta''+ \beta'^2 + \beta'(\gamma'-\alpha')].
\ear
  This effective potential was previously obtained in other notations
  for $\ep =-1$ in \cite{lecht}. A further substitution
\beq                                                       \label{to_y}
       \psi (x, t) = y(x) \e^{i\omega t}, \cm \omega = \const,
\eeq
  which is possible because the background is static, leads to the
  \Schr-like equation
\beq                                                        \label{Schr}
      y_{xx} + [\omega^2 - \Veff(x)] y =0.
\eeq
  If there is a nontrivial solution to (\ref{Schr}) with $\im \omega <0$
  satisfying some physically reasonable conditions at the ends of the range
  of $u$ (in particular, the absence of ingoing waves), then the
  static system is unstable since $\df$ can exponentially grow with $t$.
  Otherwise our static system is stable in the linear approximation. Thus,
  as usual in such studies, the stability problem is reduced to a
  boundary-value problem for \eq (\ref{Schr}) --- see, e.g., \cite
  {we99, br-hod, stepan04, stepan05, gon08, Konoplya:2005et, Konoplya:2010kv}.

  Note that all the above relations are written without fixing the background
  radial coordinate $u$.

\subsection {Gauge-invariant perturbations}

  To be sure that we are dealing with real perturbations of the static
  background rather than purely coordinate effects, it is necessary to
  construct gauge-invariant quantities.

  Small coordinate transformations $x^a \mapsto x^a + \xi^a$
  in the $(t,u)$ subspace can be written as
\beq                                                       \label{u,t trans}
    t = \ot + \Delta t (t,u), \qquad u = \ou + \Delta u (t,u),
\eeq
  where $\Delta t$ and $\Delta u$ are supposed to be small. Any scalar
  quantity with respect to such transformations, such as, e.g., $\phi (t,u)$
  acquires an increment:
\beq
      \Delta\phi = \dot\phi \Delta t + \phi'\Delta u \approx \phi'\Delta u
\eeq
  in the linear approximation since both $\dot\phi$ and $\Delta t$ are
  small. The quantity $r$, also being a scalar in the $(t,u)$ subspace
  (a 2-scalar, for short), behaves in the same way. If there are
  perturbations $\delta\phi$ and $\delta r$, the transformation (\ref{u,t
  trans}) changes them as follows:
\bearr
     \delta\phi \mapsto \overline{\delta\phi} = \delta \phi +\phi'\Delta u,
\nnn
     \delta r \mapsto \overline{\delta r} = \delta r + r' \Delta u.
\ear
  It then follows that the combination
\beq                                                          \label{inv1}
    \psi_1 \equiv  r' \delta \phi - \phi' \delta r,
\eeq
  is invariant under the transformation (\ref{u,t trans}), or
  gauge-invariant. Recall that the prime here denotes $d/du$ in the
  background static configuration.

  One can notice that combinations constructed like (\ref{inv1}) from any
  2-scalars (for example, such as $\e^\phi$ and $\beta=\ln r$, or two
  different linear combinations of $\phi$ and $r$) are also gauge-invariant.
  Moreover, gauge-invariant is $\psi_1$ multiplied by any 2-scalar or any
  combination of background quantities which are known and fixed functions
  of $u$.

  The physical properties of perturbations must not depend on which
  gauge-invariant quantity $\psi$ is chosen to describe them. Meanwhile,
  with different $\psi$, the effective potentials will, in general, also be
  different. However, given a specific background configuration, in order
  that the theory be consistent, these different potentials should lead to
  the same perturbation spectrum.

  Due to gauge invariance of $\psi_1$, equations that govern it may be
  written in any admissible gauge, in particular, the gauge $\delta\beta=0$,
  and \eq (\ref{wave}) for $\psi$ then may be considered as a result
  of substituting $\psi = (r/r')\psi_1 = r \delta \phi$ in a manifestly
  correct equation for $\psi_1 = r'\df$.

  On the similar problem regarding cosmological perturbations and
  the definition of the corresponding gauge invariants, see, e.g., the
  review \cite{brand92}.

\subsection {Regularized potential near a throat}

  The gauge $\delta\beta = 0$ (the same as $\delta r =0$) is suitable for
  describing the perturbations at any points except those where $r' =0$:
  these are throats and other critical points of $r(u)$. Indeed \cite{we99},
  putting $\delta r =0$, we forbid perturbations of the throat radius, while
  there is no physical reason for that. Technically, this restriction
  manifests itself in a generically infinite value of the potential $U(u)$
  in \eq (\ref{eq-df}) and consequently in $\Veff$ involved in the wave
  equation (\ref{wave}). Throats are only possible in the case $\ep = -1$,
  and, provided $U/r^2 < 1$ at such a throat ($u = \uth$), the potential has
  there a wall of infinite height, with the generic behavior $\Veff \sim
  1/(u - \uth)^2$ near the throat since we have there generically $r'(u)
  \sim u-\uth$. As a result, perturbations are actually independent at
  different sides of the throat, necessarily turn to zero at the throat
  itself, and we thus partly lose information on their possible properties.
  Such an incomplete treatment has led to a conclusion that anti-Fisher
  \whs\ \cite{h_ellis, br73} (that is, \wh\ solutions to \eqs (\ref{SE}),
  (\ref{EE}) with $V \equiv 0$ and $\ep = -1$) were stable under \sph\
  perturbations. A similar conclusion was made in \cite{we99} concerning
  such \whs\ and cold \bhs\ by using another (harmonic) gauge, $\delta\alpha
  = 2\delta\beta + \delta\gamma$, which does not lead to a pole in the
  effective potential, but, as follows from our further consideration in
  this paper, this analysis was also incomplete.

  It could seem that the above difficulty only concerns the gauge
  $\delta\beta =0$. However, due to the gauge-invariant nature of \eq
  (\ref{wave}) (to be verified below), it is clear that the problem is
  inherent to the background geometry itself, and the pole in the effective
  potential always emerges at a throat, if any.

  A way of avoiding the restriction $\db (\uth) =0$ is connected
  with the so-called S-deformations of the potential $\Veff$. This method
  was used in \cite{kod03, kod11} for transforming a partly negative
  potential to a positive-definite one in master equations for perturbations
  of higher-dimensional \bhs.
  Using this method, Gonzalez et al. \cite{gon08} transformed a singular
  potential to a nonsingular one for perturbations of the anti-Fisher \whs\
  and discovered the existence of an exponentially growing mode, showing
  that such \whs\ are unstable. We will try to formulate a similar scheme
  suitable for the more general field system (\ref{L}).

  Consider a wave equation of the type (\ref{wave})
\beq                                                        \label{wave1}
        \ddot \psi - \psi_{xx} + W(x) \psi = 0,
\eeq
  with an arbitrary potential $W(x)$ (whose specific example is the above
  potential $\Veff$). If there is a function $S(x)$ such that $W(x)$ is
  presented in the form
\beq                                                        \label{W(S)}
        W(x) = S^2(x) + S_x,
\eeq
  then \eq (\ref{wave1}) is rewritten as follows:
\beq                                                        \label{wave2}
        \ddot\psi + (\d_x + S)(-\d_x + S)\psi =0.
\eeq
  Now, if we introduce the new function
\beq                                                        \label{chi}
        \chi = (-\d_x + S)\psi,
\eeq
  then, applying the operator $-\d_x + S$ to the left-hand side of \eq
  (\ref{wave2}), we obtain the following wave equation for $\chi$:
\beq                                                        \label{wave3}
        \ddot\chi - \chi_{xx} + \Wreg (x)\chi = 0,
\eeq
  with the new effective potential
\beq                                                        \label{W1}
        \Wreg (x) = -S_x + S^2 = - W(x) + 2 S^2.
\eeq

  If a static solution $\psi_s(x)$ of \eq (\ref{wave1}) is known, so that
  $\psi_{s,xx} = W(x)\psi_s$, then we can choose
\beq                                                       \label{S-sol}
        S(x) = \psi_{s,x}/\psi_s
\eeq
  to carry out the above transformation.

  Generically, the function $U$ in (\ref{eq-df}) and hence the potential
  (\ref{Veff}) behave near a throat as $r'{}^{-2} \sim (u-\uth)^{-2}
  \sim x^{-2}$, where, without loss of generality, we put $x = 0$ at the
  throat. Assuming that the potential $W(x)$ behaves in such a way, let us
  look if a transition to $\Wreg$ can really remove this singularity. Above
  all, we see that according to (\ref{W1}), such removal is only possible if
  $W \to +\infty$ as $x\to 0$ since we are dealing with real quantities.
  Thus a potential wall in $W(x)$ can be removed but a potential well
  cannot.

  A positive pole $W \sim x^{-2}$ can be removed in $\Wreg$ if
\beq                                                             \label{S0}
    S \approx  - 1/x   \ \ \then \ \  \psi_s \propto 1/x.
\eeq
  It is a necessary condition for regularizing the potential. Besides, to
  avoid a singularity of $\Wreg$, $\psi_s$ must be nonzero in the whole
  range of $u$. Moreover, according to (\ref{W1}), it is clear that near the
  throat $x=0$ in this case
\beq                                                          \label{W2}
      W (x) \approx  2/x^2.
\eeq
  Thus, to be regularized by the procedure described, the pole in $W(x)$
  must behave as (\ref{W2}). Let us show that this is generically the case
  for the potential $W = \Veff (x)$ given by \eq (\ref{Veff}).

  Suppose such a generic situation, so that

\medskip\noi
  (i) the function $\beta(x)$ in the background metric is expanded
  near its minimum (the throat) in powers of $x$ as follows:
\beq                                                         \label{b(x)}
       \beta(x) = \beta_0 + \half \beta_2 x^2
                + \fract{1}{6}\beta_3 x^3 + \ldots.
\eeq
  where $\beta_{0,2,3}$ are constants;

\medskip\noi
  (ii) the background quantity $\phi'(x) \ne 0$ at $x=0$.

\medskip
  Here and till the end of the section, we use the coordinate freedom to
  choose the ``tortoise'' radial coordinate $u=x$ specified by the condition
  $\alpha = \gamma$. All functions are considered as power series in $x$ at
  small $x$.

  Let us estimate $\Veff$ (\ref{Veff}).
  The term that determines the pole at $x=0$ is
\beq
       W_{{\rm pole}} (x)=                                    \label{Wpole}
       \e^{2\gamma} (V - \e^{-2\beta})\frac{\phi'^2}{\beta'^2},
\eeq
  where we have put $\ep = -1$ since throats are possible only in this case.
  By (\ref{b(x)}),
\beq                                                         \label{beta'}
    \beta_x = \beta_2 x + \half \beta_3 x^2 + \ldots.
\eeq
  Now we use the equations governing the static configuration: from
  (\ref{22-0}) it follows that $\e^{2\gamma} (\e^{-2\beta} - V) = \beta_{xx}$
  at $x=0$, and then from (\ref{G00-0}) we find that $\phi_x^2 = 2\beta_{xx}$
  at $x=0$. Thus from assumption (ii) it follows $\beta_2 \ne 0$.
  Substituting all this to (\ref{Wpole}), we find that it behaves precisely
  as required in (\ref{W2}).

  Thus we have shown that, for a (generic) throat in a solution to
  \eqs (\ref{e-phi0})--(\ref{G00-0}), the effective potential $\Veff$ for
  \sph\ perturbations satisfies the necessary condition for regularization
  by the above method.

  Whether or not this regularization really works and leads to a regular
  boundary-value problem for the perturbations, should be investigated for
  specific background configurations. (In particular, one should also take
  into account the other singular term in the potential $\Veff$,
  proportional to $V_{\phi}/\beta_x$: the terms $\propto 1/x$ evidently
  depend on the finite part of $S(x)$.)

  A positive example of such a study, concerning anti-Fisher \whs, is known
  from \cite{gon08}; we here show that the other two branches of the
  anti-Fisher solution are also such examples.

\subsection {Regular perturbations near a throat}

  Suppose we have found a solution $\chi(x,t)$ to \eq (\ref{wave3}),
  satisfying the appropriate boundary conditions. The function $\chi$ is
  regular at $x = 0$ since the potential $\Wreg$ is regular there. If $\chi$
  is a growing function of $t$, it probably indicates an instability of the
  initial static configuration; but it is indeed the case only if this
  $\chi(x,t)$ creates regular perturbations of the metric functions
  $\alpha$, $\beta$, $\gamma$ and the scalar field $\phi$.

  Let us look how it happens. Given $\chi(x,t)$, a solution to (\ref{wave1}),
  or (\ref{wave}), is found as $\psi = (\d_x + S)\chi$. Generically, $\chi$
  is finite at $x = 0$ while $S \approx 1/x$ at small $x$, hence $\psi \sim
  1/x$, and according to (\ref{to_psi}) we obtain $\delta\phi \to \infty$ at
  the throat. This result is in fact quite natural since the relation
  (\ref{to_psi}) corresponds to the gauge $\delta r = 0$, in which the
  throat radius is fixed, whereas we were seeking perturbations with nonzero
  $\delta r$ on the throat. So it is necessary to pass on to another gauge,
  which is easily done due to gauge invariance of the quantity $\psi$ given
  by
\beq                                                         \label{psi+}
        \psi = r \df - \frac{r\phi_x}{r_x} \delta r.
\eeq
  Namely, a finite expression for $\delta r$ is obtained in the gauge
  $\delta\phi = 0$ provided $\phi_x(0) \ne 0$ since then
\beq                                                          \label{dr}
        \delta r = - \frac{r_x}{r\phi_x} \psi,
\eeq
  while the product $r_x\psi$ is finite. It remains to find $\da$ and $\dg$
  from the perturbation equations in the gauge $\df =0$. From \eqs
  (\ref{01-1}) and (\ref{e-phi1}) we find
\bearr                                                       \label{da-dr}
    \beta_x \da = \db_x + \db (\beta_x -\gamma_x),
\yyy                                                         \label{dg-dr}
    \dg_x = \da_x - 2\db_x- \frac{2\ep}{\phi_x} V_{\phi}\e^{2\alpha}\da,
\ear
  but here we are again facing a problem: according to (\ref{da-dr}), in
  general $\da$ diverges at the throat where $\beta_x=0$. This divergence is
  only avoided if the right-hand side of (\ref{da-dr}) behaves like
  $\beta_x \sim u - \uth \sim x$.

  Surprisingly, it is the case in a generic situation, as can be verified in
  a general form using near-throat expansions. Indeed, let us preserve the
  above assumptions (i) and (ii) and assume, in addition, that

\medskip\noi
  (iii) the function $\chi(x,t)$ that solves \eq (\ref{wave3}) is finite and
  nonzero at $x=0$,

\medskip\noi
  (iv) the function $S(x)$ behaves at small $x$ according to (\ref{S0}).

\medskip

  Our task is to estimate the right-hand side of \eq (\ref{da-dr}) in the
  order $O(x^0)$: if it is zero, it means that $\beta_x \da \sim x \sim
  \beta_x$ and thus $\da(0)$ is finite; the remaining metric perturbation
  $\dg$ is then found from (\ref{dg-dr}) and is also finite.

  Taking $\delta\beta = \delta r/r$ from (\ref{dr}) and substituting $\psi$
  as $\psi = \chi_x + S\chi \sim -\chi_0/x$, where $\chi_0 =\chi(0)$, we find
\[
    \db = \frac{\chi_0\e^{-\beta}}{\phi_x}
            \Big(\beta_2 + \half \beta_3 x +\ldots\Big).
\]
  Then we substitute this expression to (\ref{da-dr}) to obtain
\[
     \beta_x \da \approx \frac{\chi_0\e^{-\beta}}{\phi_x^2}
        \Big[ -\phi_{xx}\beta_2
            + \half \phi_x (\beta_3 - 2\gamma_x \beta_2)\Big],
\]
  where all quantities are taken at $x=0$. Now, $\phi_{xx}$ can be expressed
  from the background equation (\ref{e-phi0}), $\beta_{xx}$ from
  (\ref{22-0}); we can use the fact that $\beta_2 = \beta_{xx}(0)$ etc., and
  we can also ignore all terms proportional to $\beta_x$. After these
  substitutions we finally obtain
\[
     \beta_x \da \propto \e^{2\gamma} V_\phi
                \Big [\ep \beta_{xx} + \half \phi_x^2 \Big]_{x=0}.
\]
  But the expression in the square brackets vanishes due to the difference
  of \eqs (\ref{00-0}) and (\ref{11-0}), which proves that $\beta_x \da =
  O(x)$ and thus $\da(0)$ is finite.

  We conclude that {\it under the generic assumptions (i)--(iv),
  regularization of the potential $\Veff$ always leads to finite
  perturbations of the background static solution.}

\section {Instabilities of the Fisher and anti-Fisher solutions}

\subsection {The static solutions}

  Let us recall the well-known \ssph\ solutions to the field equations
  (\ref{SE}) and (\ref{EE}) for zero potential, $V \equiv 0$. In the case
  $\eps = +1$ this solutions was found by I.Z. Fisher in 1948 \cite{fisher}
  and afterwards repeatedly re-discovered. For $\eps = -1$ the corresponding
  solution was first obtained, to our knowledge, by Bergmann and Leipnik
  \cite{Afish} and also repeatedly re-discovered. But these authors used
  the curvature coordinates [i.e., the condition $u\equiv r$ in terms of the
  metric (\ref{ds})], which are not well suited for the problem, and maybe
  therefore they did not give a clear interpretation of the solutions.

  The solution can be written jointly for $\ep = \pm 1$ if one uses the
  harmonic coordinate $u=v$ in the metric (\ref{ds}), corresponding to the
  coordinate condition $\alpha(v) = \gamma(v) + 2\beta(v)$ \cite{br73}.
  The scheme of obtaining it is as follows. In these coordinates, the scalar
  field equation and the ${0\choose 0}$ component of (\ref{EE}) read simply
  $\phi''=0$ and $\gamma''=0$, respectively (the prime here stands for
  $d/dv$), so that, choosing the zero point of $\phi$ and the time scale,
  without loss of generality we can write $\phi = Cv$ and $\gamma = -mv$
  with $C, m = \const$. Furthermore, a sum of the components ${0\choose 0}$
  and ${2\choose 2}$ of (\ref{EE}) leads to the Liouville equation
  $(\beta + \gamma)'' = \e^{2\beta+ 2\gamma}$ whose solution is
\beq                                                       \label{def_s}
      \e^{-\beta-\gamma} = s(k,v) := \vars     {
                        k^{-1}\sinh kv,  \ & k > 0 \\
                                    v,  \ & k = 0 \\
                        k^{-1}\sin kv,   \ & k < 0.  }
\eeq
  As a result, the solution as a whole reads
\bear                                              \label{fish}
     ds^2 \eql \e^{-2mv}dt^2 - \frac{\e^{2mv}}{s^2(k,v)}
                 \biggr[\frac{dv^2}{s^2(k,v)} + d\Omega^2\biggl],
\cm
     \phi = Cv,
\ear
  where the integration constants $m$ (the Schwarzschild mass), $C$
  (the scalar charge) and $k$ are related by the equality
\beq                                                       \label{int_0}
            2k^2\sign k = 2m^2 + \ep C^2,
\eeq
  obtained after substituting all found functions to the Einstein equation
  $G^1_1 = \ldots$ which does not contain second-order derivatives in $v$
  and is an integral of other equations.

  The coordinate $v$ is defined in the whole range $v > 0$ for $k\geq 0$
  and in the range $0 < v < \pi/|k|$ for $k < 0$. The value $v=0$ in all
  cases corresponds to flat spatial infinity, so that at small $v$ the
  spherical radius is $r(v) \approx 1/v$, and the metric becomes
  approximately Schwarzschild.

  In the case $k > 0$, it is helpful to pass over to the quasiglobal
  coordinate coordinate $u$ (defined by the condition $\alpha + \gamma=0$
  in (\ref{ds})) by the transformation
\beq
         \e^{-2kv} = 1 - 2k/u =: P(u),                      \label{def_P}
\eeq
  and the solution takes the form
\bear                                                       \label{ds+}
     ds^2 = P^{a} dt^2 - P^{-a} du^2
                                 - P^{1 - a} u^2 d\Omega^2,
    \cm
        \phi = -\frac{C}{2k} \ln P(u),
\ear
  with the constants related by
\beq                                                        \label{int_0+}
    a = m/k, \cm  a^2 = 1 -\ep C^2/(2k^2).
\eeq

  {\bf The Fisher solution} \cite{fisher} corresponds to $\eps = +1$, hence
  according to (\ref{int_0}), it consists of a single branch $k > 0$
  and, in (\ref{ds+}), $|a| < 1$. It is defined in the range $u > 2k$,
  and $u = 2k$ is a naked central ($r=0$) singularity which is attractive
  for $m >0$ and repulsive for $m < 0$. The Schwarzschild solution is
  restored at $C=0$, $a = 1$ for $m > 0$ and at $C=0$, $a=-1$ for $m < 0$.

  The solution for $\eps = -1$ (that is, for a phantom scalar field) is
  conveniently termed {\bf the anti-Fisher solution}, by analogy with de
  Sitter and anti-de Sitter. According to three variants of the function
  (\ref{def_s}), this solution splits into three branches with the
  following properties.

\medskip\noi
  {\bf Branch A}, $k > 0$: the solution again has the form (\ref{ds+}), but
  now $|a| > 1$.  For $m <0$, that is, $a <-1$, we have, just as in the
  Fisher solution, a repulsive central singularity at $u = 2k$. The
  situation is, however, drastically different for $a > 1$.

  Indeed, the spherical radius $r$ in all such cases has a finite minimum
  at $u = u_{\rm th} = (a+1)k$, corresponding to a throat of the
  size
\beq                                                        \label{r_th}
    r(u_{\rm th}) = r_{\rm th} = k (a+1)^{(a+1)/2} (a-1)^{(1-a)/2},
\eeq
  and tends to infinity as $u\to 2k$. Moreover, for $a=2,\ 3,\ \ldots$
  the metric exhibits a horizon of order $a$ at $u = 2k$ and admits a
  continuation to smaller $u$ \cite{we-afish}. A peculiarity of such horizons
  is their infinite area. Such \asflat\ configurations with horizons of
  infinite area have been termed cold \bhs\ (CBHs) \cite{we99} since all of
  them have zero Hawking temperature. The throat radius (\ref{r_th}) does
  not coincide with the Schwarzschild mass $m = ak$ (in usual units, half the
  Schwarzschild radius $2Gm/c^2$) but is of the same order of magnitude.

  Furthermore, it can be verified using (\ref{k_i}) \cite{we-afish} that the
  metric (\ref{ds+}) has a curvature singularity at $u=2k$ if $a < 2$
  (except for the Schwarzschild case $a=1$), a finite curvature if $a=2$ and
  zero curvature if $a > 2$.

  For non-integer $a > 2$, the qualitative behavior of the metric as
  $r\to 2k$ is the same as near a horizon of infinite area, but a
  continuation beyond it is impossible due to non-analyticity of the
  function $P^a(u)$ at $u = 2k$. Since geodesics terminate there at a finite
  value of the affine parameter, this is a space-time singularity (a {\it
  singular horizon\/} as it is named in \cite{we99}) even though the
  curvature invariants tend there to zero.

\medskip\noi
  {\bf Branch B}, $k = 0$: the solution is defined in the range $u \in \R_+$
  and is rewritten in terms of the quasiglobal coordinate $u = 1/v$ as
  follows:
\bear
       ds^2 = \e^{-2m/u}dt^2                                \label{ds_0}
                - \e^{2m/u}[du^2 + u^2 d\Omega^2],
\cm
       \phi = C/u.
\ear
  As before, $u = \infty$ is a flat infinity, while at the other extreme,
  $u\to 0$, the behavior is different for positive and negative mass. Thus,
  for $m < 0$, $u =0$ is a singular center ($r=0$), while for $m > 0$, $r
  \to \infty$ and all $K_i \to 0$ as $u \to 0$. This is again a singular
  horizon: despite the vanishing curvature, the non-analyticity of the
  metric in terms of $u$ makes its continuation impossible. The throat
  occurs at $u = m$ and has the size $e\cdot m$, $e$ being the base of
  natural logarithms.

\medskip\noi
  {\bf Branch C}, $k < 0$: the solution describes a \wh\ with two flat
  asymptotics at $v=0$ and $v = \pi/|k|$. The metric has the form
  \cite{h_ellis, br73}
\bear                                                            \label{ds-}
    ds^2 \eql \e^{-2mv} dt^2 - \frac{k^2\e^{2mv}}{\sin^2 (kv)}
        \biggl[ \frac{k^2\,du^2}{\sin^2 (kv)} + d\Omega^2 \biggr]
\nn
      \eql \e^{-2mv} dt^2 - \e^{2mv} [du^2 + (k^2 + u^2) d\Omega^2],
\ear
  where $v$ is expressed in terms of the quasiglobal coordinate $u$,
  defined on the whole real axis, by $\ok v =  \cot^{-1} (u/\ok)$, where
  we have denoted $-k = \ok > 0$. If $m > 0$, the \wh\ is attractive for
  ambient test matter at the first asymptotic ($u\to \infty$) and repulsive
  at the second one ($u \to - \infty$), and vice versa in case $m < 0$.  For
  $m = 0$ one obtains the simplest possible \wh\ solution, sometimes called
  the Ellis \wh, although Ellis \cite{h_ellis} actually discussed these
  solutions with any $m$.

  The \wh\ throat occurs at $u = m$ and has the size
\beq                                                           \label{thr-}
    r_{\rm th} = (m^2 + \ok^2)^{1/2}
            \exp \left(\frac{m}{\ok} \cot^{-1}\frac{m}{\ok}\right).
\eeq

\subsection{Perturbations: the Fisher solution}

  For a massless scalar field [$V \equiv 0$ in (\ref{L})], the effective
  potential in (\ref{wave}) takes the following form in terms of the
  quasiglobal coordinate $u$:
\beq                                                          \label{Veff-A}
    \Veff = -\ep \frac{A\phi'^2}{r'^2} + \frac{A r''}{r}
            + \frac{A' r'}{r}
        = -\ep \frac{A\phi'^2}{r'^2} + \frac{A}{r^2} - \frac{A^2 r'^2}{r^2},
\eeq
  where $A(u) := \e^{2\gamma(u)} = \e^{-2\alpha(u)}$, and the second
  equality in (\ref{Veff-A}) follows from \eq (\ref{22-0}).

  Calculating $\Veff$ for the solution (\ref{ds+}), we find a common
  expression for both $\ep = +1$ and $\ep =-1$:
\beq                                                          \label{Veff+}
     \Veff (u) = k P^{2a}
        \frac {2au^3-3(1+a)^2ku^2+2(3+4a+3a^2+2a^3)k^2u-(1+a)^4k^3}
            {u^2(u-k(1+a))^2(u-2k)^2}.
\eeq
  Since in the Fisher solution $a < 1$, the binomial $u-k(1+a)$ is positive
  at all $u \geq 2k$, and the only singularity in $\Veff$ is $u\to 2k$,
  coinciding with the singularity of the background solution. Near the
  singularity, at which according to (\ref{to_x}) we can put $x=0$,
  $\Veff(u) \sim - 1/(4x^2)$, a negative pole in agreement with \cite{br-hod}
  and many subsequent papers.

  The boundary condition at spatial infinity ($u\to \infty$, $x \to \infty$)
  is natural: $\df\to 0$, or $\psi\to 0$. For $u\to 2k$, where the
  background field $\phi$ tends to infinity, the boundary condition is not
  so evident. In \cite{br-hod} and other papers, dealing with minimally
  coupled or dilatonic scalar fields, the minimal requirement was used
  providing the validity of the perturbation scheme:
\beq
        |\df/\phi | < \infty.                                \label{weak}
\eeq
  (The requirement of absence of ingoing waves then does not lead to
  further restrictions.)
  Under this boundary condition it is easy to conclude that there are
  solutions to the \Schr-like equation (\ref{Schr}) with any $\omega^2 < 0$,
  which means that the static field configuration is unstable, in agreement
  with the previous work \cite{br-hod} (see also \cite{we-afish} for
  details).

\subsection{Perturbations: the anti-Fisher solution}

\subsection*{Branch A}

  The effective potential has the same form (\ref{Veff+}), but now, since
  $a > 1$ (we restrict ourselves to this case providing $m > 0$), the
  potential has a positive pole at $u = \uth = k(a+1) > 2k$, the throat in
  the background configuration.

  To remove this singularity, we were able to find the following simple
  static solution to \eq (\ref{wave}):
\bearr                                                      \label{stat+1}
    \psi_{s+}(u) \propto r(u) \frac{au-\uth}{u-\uth},
\ear
  Applying the technique described in Sec.\,2.5 with these $\psi_s(u)$,
  we the new effective potential $\Wreg = W_A(u)$:
\bear                                                    \label{regV+1}
    W_{A}(u) \eql \left(1-\frac{2k}{u}\right)^{2a}
            \frac {N (u)}{u^2 (u-2k)^2 (au- \uth)^2},
\nn
     N (u) \eql 3 (1+a)^4 k^4 - 2a (6 + 19a + 16a^2 + 3a^3) k^3 u
\nnn \cm
          + 3 a^2 (9 + 10a + a^2) k^2 u^2 - 2a^2 (4 + 5a)k u^3+ 2a^2 u^4.
\ear
\begin{figure*} 
\centering
\resizebox{.45\linewidth}{!}{\includegraphics*{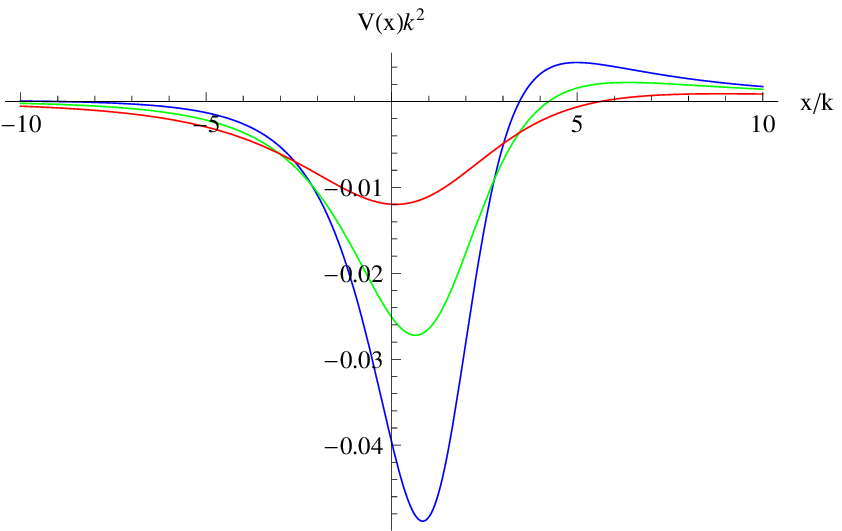}}
\resizebox{.45\linewidth}{!}{\includegraphics*{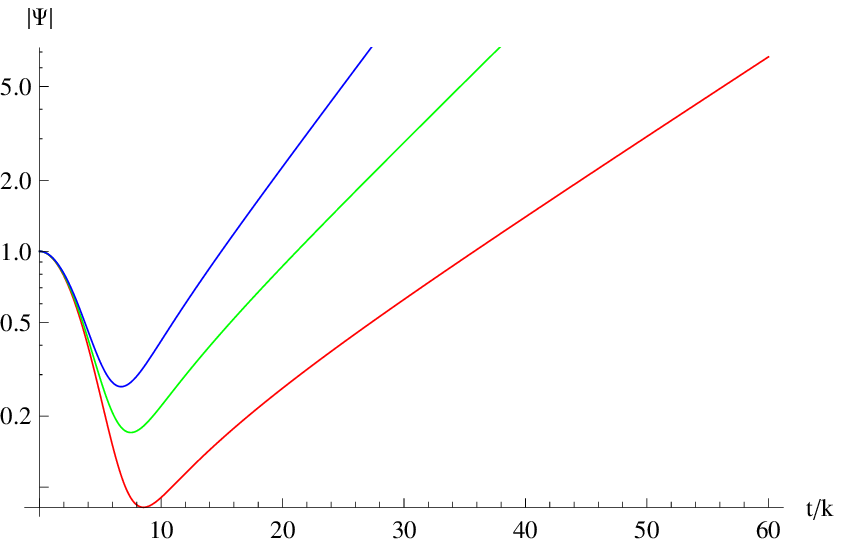}}
\caption{The regularized effective potential $W_A$ (left panel) and
    the time-domain profiles (right panel) for the Branch A solutions
    with $a=3/2$ (blue), $a=2$ (green), $a=3$ (red).
    Smaller values of $a$ correspond to deeper potential wells and
    a more rapid growth of perturbations.}   \label{bAfig}
\end{figure*}

  The potential (\ref{regV+1}) has no singularities at $r > 2k$ and is
  partly negative, which should in general lead to an instability. To prove
  it, one can use the method of time-domain integration \cite{Gund93}
  allowing for following the time evolution of the perturbations under
  prescribed initial and boundary conditions. The latter, as usual, must
  provide the absence of ingoing waves from the boundary, and in our case it
  is sufficient to simply require $\psi \to 0$ and $\chi \to 0$ as $x\to \pm
  \infty$ for all three branches of the anti-Fisher solution. The reason is
  that the effective potential $\Veff$ as well as the regularized potentials
  vanish at large $|x|$, and therefore modes of interest, those with
  $\omega^2 < 0$, should exponentially decay at large $|x|$.

  Examples of plots for the potential (\ref{regV+1}) and the results of
  time-domain integration are shown in Fig.\,\ref{bAfig}. By fitting of the
  profile we find that the perturbations grow approximately as $\psi \propto
  e^{0.25t/m}$.

\begin{figure*}         
\centering
\resizebox{.45\linewidth}{!}{\includegraphics*{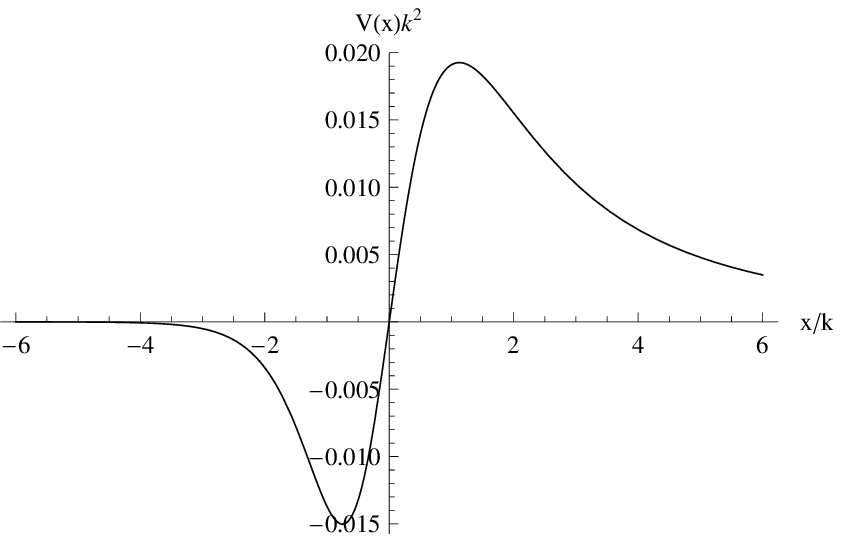}}
\resizebox{.45\linewidth}{!}{\includegraphics*{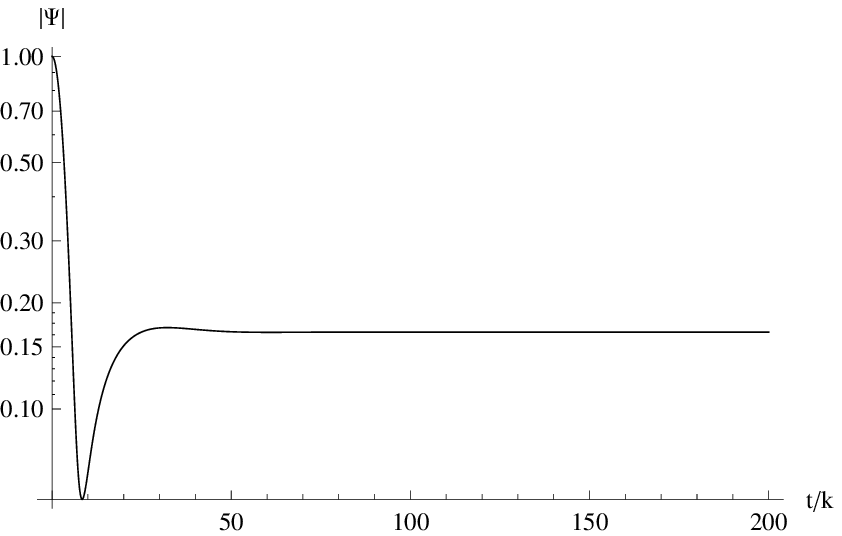}}
\caption{The potential $W_A$ (left panel) and the time-domain profile
    (right panel) for the Branch A solution with $a=1$.}     \label{bA1fig}
\end{figure*}

  In the limit $a\to 1$ the regularized potential still has a negative gap,
  however, at $a=1$ no growing mode is observed, and a stationary solution
  dominates at late times (Fig.\,\ref{bA1fig}), while at any finite $a-1$
  the perturbation does grow. This shows how the instability is
  ``dying out'' when approaching the Schwarzschild solution.
  One can recall that in the genuine Schwarzschild case there is no
  scalar field, and the modes we are considering here simply do not exist.

\subsection*{Branch B}

  For the solution (\ref{ds_0}), the effective potential (\ref{Veff-A})
\beq
    \Veff(u) = m\exp\left(-\frac{4m}{u}\right)          \label{Veff-0}
    \frac{2 u^3 - 3 mu^2 + 4m^2 u - m^3}{(u - m)^2 u^4}
\eeq
  is singular at $\uth = m$. We again find a static solution to (\ref{wave})
\beq
    \psi_{s0}(u) \propto \frac{u^2\,\e^{m/u}}{u-m}        \label{stat0}
\eeq
  and perform the transformation described in Sec.\,2.5.

\begin{figure*}                
\centering
\centerline{\resizebox{.45\linewidth}{!}{\includegraphics*{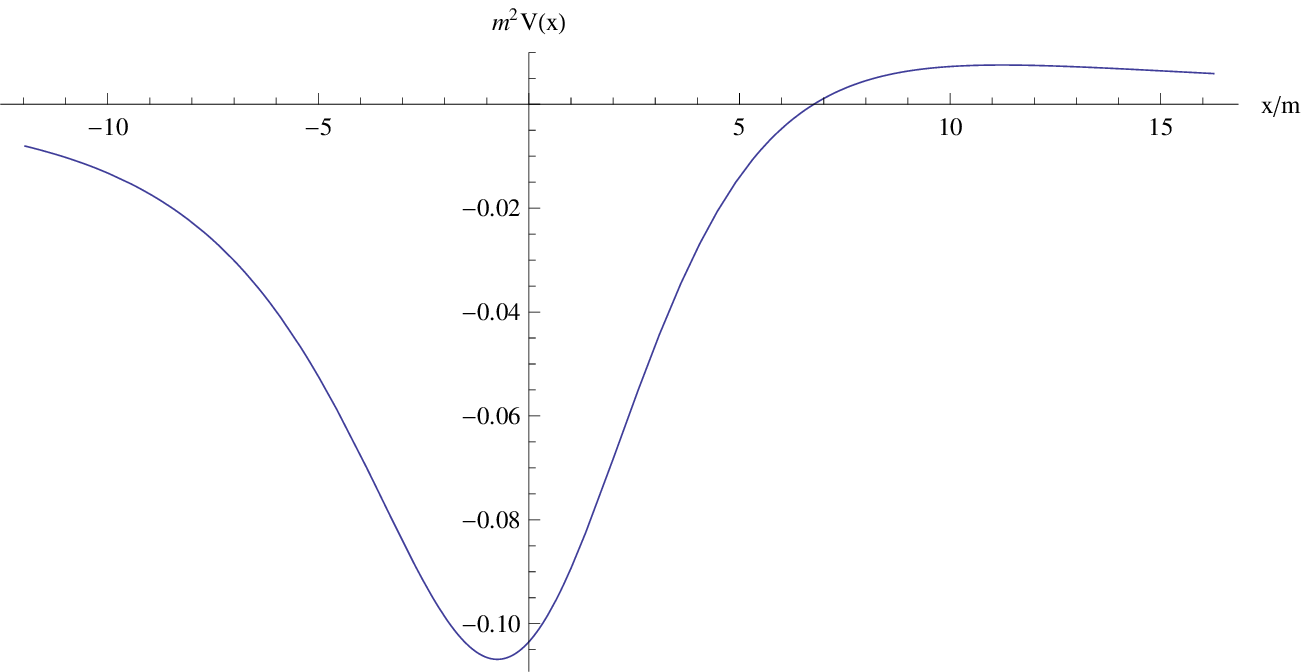}}
\resizebox{.45\linewidth}{!}{\includegraphics*{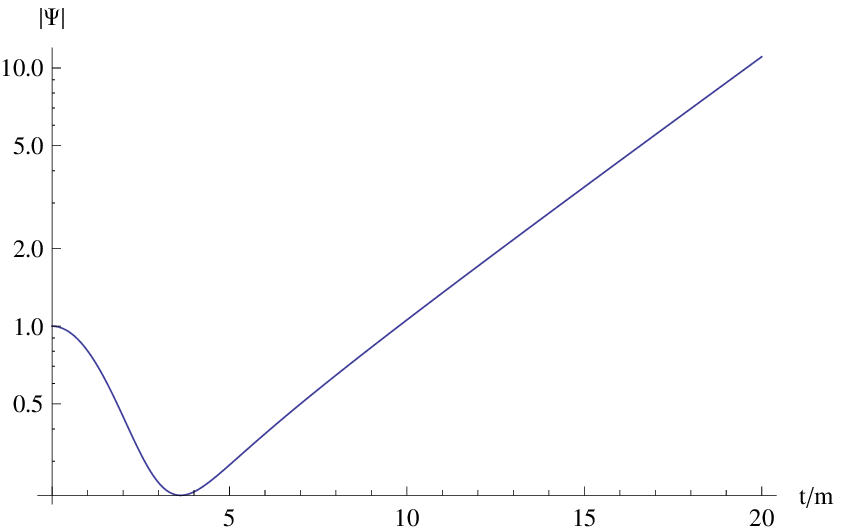}}}
\caption{The potential $W_B$ (left panel) and the time-domain profile
    (right panel) for the Branch B solution.}\label{bBfig}
\end{figure*}

  The regularized effective potential has the form of an inverse potential
  barrier:
\beq
    W_B (u) = \exp\left(-\frac{4m}{u}\right)        \label{regV0}
        \frac{3 m^2 - 10 m u + 2 u^2}{u^4},
\eeq
  The perturbations grow approximately as $\psi \propto e^{0.23t/m}$
  (see an example in Fig.\,\ref{bBfig}).

\subsection*{Branch C}

\begin{figure*}                       
\centering
\resizebox{.45\linewidth}{!}{\includegraphics*{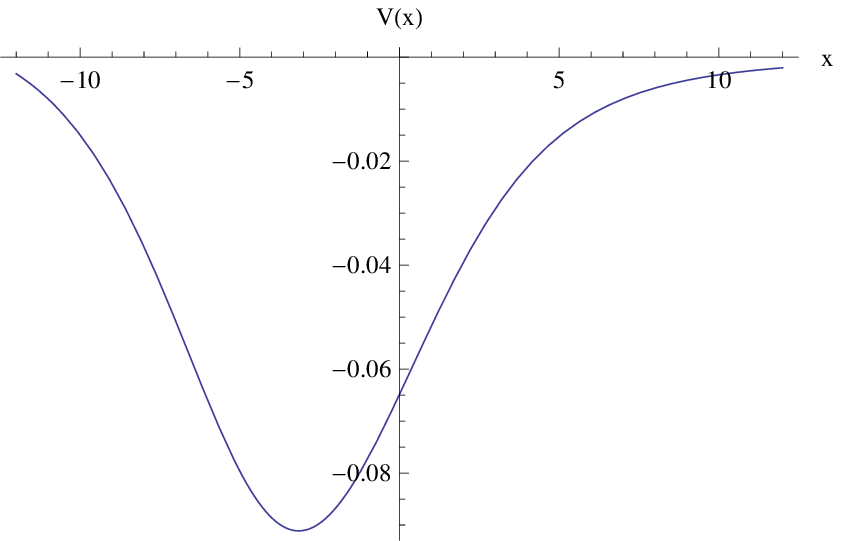}}
\resizebox{.45\linewidth}{!}{\includegraphics*{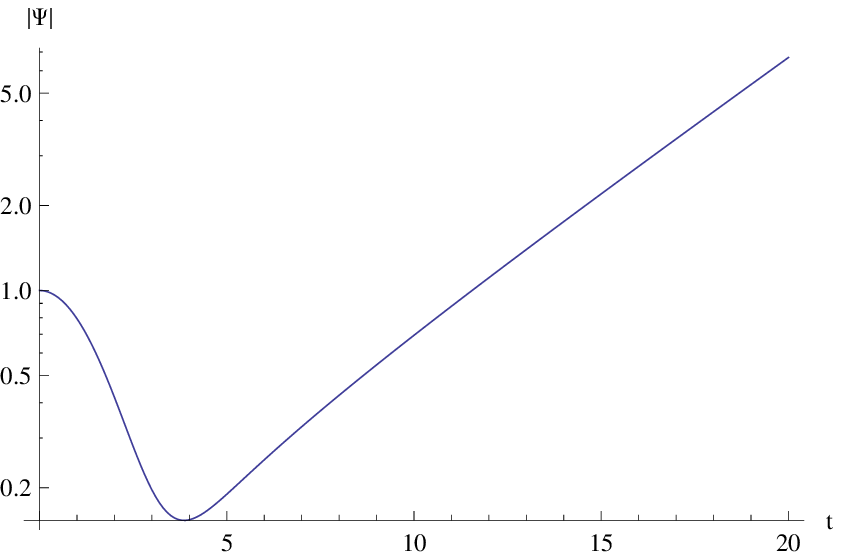}}
\caption{The potential $W_{C1}$ (left panel) and the time-domain profile
        (right panel) for Branch C ($k=m=1$).}  \label{bC1fig}
\end{figure*}

  For the wormhole solution (\ref{ds-}), the potential (\ref{Veff-A}) has the
  form
\beq                                                          \label{Veff-}
   \Veff(u) = \e^{-4m v}
    \frac{2k^4 + k^2 (3m^2 - 2mu + 3 u^2) -m(m^3 -4m^2 u +3m u^2 -2u^3)}
        {(m - u)^2 (k^2 + u^2)^2}.
\eeq
  It again exhibits a positive pole on the throat. To regularize it, we can
  take any of the two static solutions
\bearr
     \psi_{s-1}(u)\propto                                      \label{stat-1}
         \frac{r(u)}{u-m}[k^2 + m^2 -mv (k^2+mu)],
\yyy
      \psi_{s-2}(u)\propto                                     \label{stat-2}
         \frac{r(u)}{u-m} (k^2 + mu).
\ear
  The first of them coincides with the solution found by Gonz\'alez et al.
  in \cite{gon08}. As a result, we reproduce their regular potential
\beq                            \label{WC1}
    W_{C1}(u) = e^{-4 m v}\frac{N_1(u)}{(k^2 +
                u^2)^2 (k^2 (m v-1) + m^2 (u v-1))^2},
\eeq
  where
\bearr
    N_1(u) = -3 k^6 (m v-1)^2 + m^4 (3 m^2 - 10 m u + 2 u^2) (u v-1)^2
\nnn
        + k^2 m^3 (u v-1) (8 u + 2 m^2 v + m (5 - 15 u v)) +
            k^4 m (m v-1) (2 u + m^2 v + m (9 - 12 u v)).
\earn
  A further investigation revealing an unstable mode of perturbations
  is described in detail in \cite{gon08} (Fig.\,\ref{bC1fig}).

\begin{figure*}                              
\centering
\resizebox{.45\linewidth}{!}{\includegraphics*{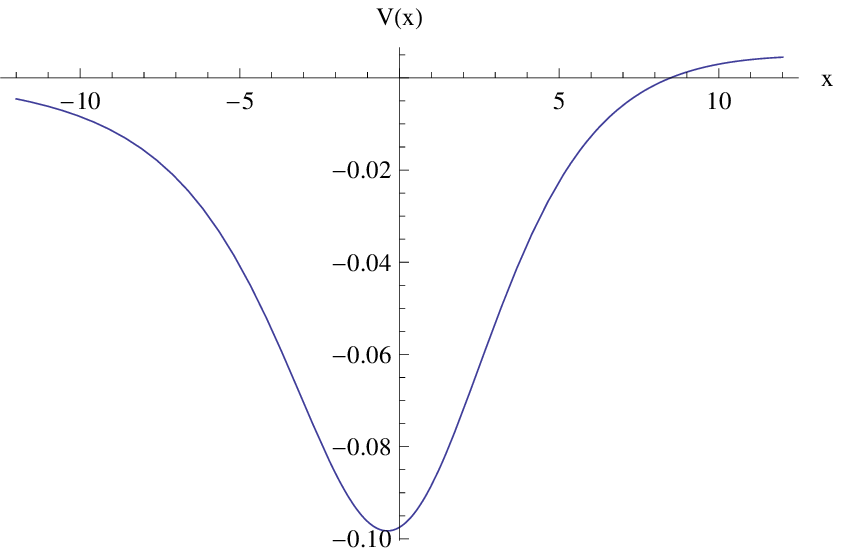}}
\resizebox{.45\linewidth}{!}{\includegraphics*{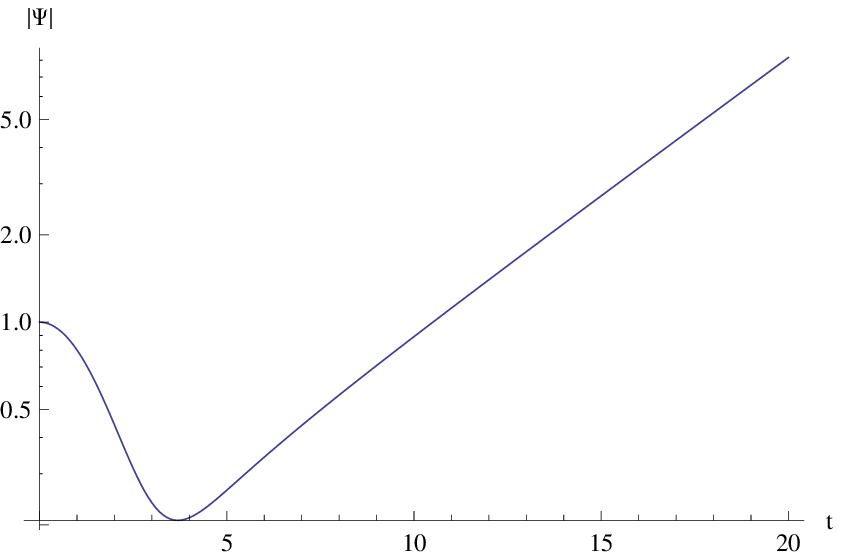}}
\caption{The potential $W_{C2}$ (left panel) and the time-domain profile
        (right panel) for Branch C ($k=m=1$).}  \label{bC2fig}
\end{figure*}

  The second static solution (\ref{stat-2}) can be used to demonstrate that,
  despite the different regularized potentials, the physical result, namely,
  the perturbation growth rate remains the same. Indeed, the regularized
  potential produces by the solution (\ref{stat-2}) reads
\beq                        \label{WC2}
    W_{C2}(u)=e^{-4 mv} \frac{-3 k^6 + k^4 m (m - 12u) + k^2 m^2 (2m - 15u)u
       + m^2 u^2 (3 m^2 - 10 m u + 2 u^2))}{(k^2 + m u)^2 (k^2 + u^2)^2},
\eeq
  and leads to perturbations growing as shown in Fig.\,\ref{bC2fig}, at the
  same rate as in Fig.\,(\ref{bC1fig}).

\section {Concluding remarks}

  We have demonstrated the instability of all \afi\ solutions to the
  Einstein-scalar equations under \sph\ perturbations, thus confirming and
  extending the conclusions of \cite{gon08} made for Branch C solutions
  (\whs). It turns out that in almost all cases the characteristic time of
  perturbation growth is of the order of the time needed for a light signal
  to cover a distance equal to the throat radius.

  We have found out that the S-deformation method of regularizing the
  effective potential for the perturbations, having a positive pole at a
  throat (a minimum of the spherical radius $r(u)$), used in \cite{gon08},
  is applicable to any \ssph\ configurations of self-gravitating scalar
  fields with self-interaction potentials. We have shown in a general form
  that, under some generic assumptions [items (i)--(iv) in Sections 2.5 and
  2.6], the potential $\Veff$ near the throat has the form admitting
  regularization, and that a regular mode found as a solution to the
  regularized equation leads to a regular perturbation of the initial
  configuration. The latter circumstance is quite non-trivial because
  regularity is required not only for solutions to the master equation but
  also for all metric coefficients.

  However, application of this methodology to specific solutions with
  self-interaction potentials $V(\phi) \ne 0$ faces the problem of
  explicitly finding the function $S(x)$ satisfying \eq (\ref{W(S)}), or,
  equivalently, a proper static solution to the master equation (a zero
  mode). We hope to extend this study to some configurations with nonzero
  $V(\phi)$ in the near future.

  One more important extension of the presently obtained results should be
  mentioned: \eqs (\ref{SE}), (\ref{EE}) are well known to be not only the
  equations of general relativity but also the Einstein-frame equations of
  the general scalar-tensor and $f(R)$ theories of gravity, see, e.g.,
  \cite{wagoner, maeda89, nood03, meng03}. Note that in $f(R)$ theories, the
  Einstein frame always contains a scalar field with nonzero $V(\phi)$. The
  observable physics in such theories is usually described in the
  corresponding Jordan conformal frame, whose metric differs from that of
  the Einstein frame only by a conformal factor that varies from theory to
  theory. A transition from one frame to the other is, from the viewpoint
  of differential equations, simply a substitution, therefore the stability
  study can safely be performed in the Einstein frame. Then, if the
  conformal factor is everywhere regular and nonzero (including the end
  points of the coordinate range), the conformal mapping preserves the
  boundary conditions for regular perturbations, and the stability
  conclusions obtained in the framework of general relativity are readily
  extended to the corresponding solutions of these generalized theories. In
  particular, we can assert that all vacuum \ssph\ solutions in
  scalar-tensor theories, connected with the (anti-)Fisher solutions by
  everywhere regular conformal factors, are unstable.

  However, in many cases the conformal factors bear nontrivial features,
  i.e., somewhere blow up or vanish, and this can affect the boundary
  conditions for the perturbation equations; in any such case a separate
  study is necessary. Consider, for instance, the counterpart of Fisher's
  solution (\ref{ds+}) in the Brans-Dicke (BD) scalar-tensor theory, where
  the Jordan-frame metric is $g^{\rm J}\mn = (1/\Phi) g\mn$; $\Phi = \exp
  (\phi/\sqrt{\omega + 3/2})$ is the BD scalar field, $\omega > -3/2$ is the
  BD coupling constant, and $|a| < 1$. It is the so-called Brans class 1
  solution. In both frames, the value $u = 2k$ is a naked singularity (see
  more details on these solutions in \cite{we99, skvo10}). In the stability
  study, to formulate a boundary condition at this singularity, for Fisher's
  solution we have used the minimal requirement (\ref{weak}),
  $|\df/\phi| < \infty$, providing the validity of the perturbation scheme,
  and we then concluded that the background solution is unstable. In the BD
  picture, it is more reasonable to require that the perturbed conformal
  factor $1/\Phi$ behave not worse than the unperturbed one, i.e., $|\delta
  \Phi/\Phi| < \infty$. However, since $\delta\Phi/\Phi \sim\df$, we arrive
  at the condition $|\df| < \infty$ which is more restrictive then
  (\ref{weak}) if $\phi \to \infty$, and this made us conclude in
  \cite{we99} that this BD solution is stable.

  Another example of using the conformal mapping between Jordan and Einstein
  frames for stability studies can be found in \cite{stepan04, stepan05},
  where the instability was proved for electrically neutral and charged
  \whs\ supported by nonminimally (in particular, conformally) coupled
  scalar fields \cite{br73, stepan05, bviss00}. In this case, there is a
  drastic difference between the manifold structures in the two conformal
  frames: in the Einstein frame, without an electric charge, it is the
  Fisher solution with the metric is $g\mn$ that has a singularity at
  $u=2k$. In Jordan's, this singularity is removed due to the conformal
  factor, the solution is continued beyond this (now regular) sphere and has
  a flat asymptotic at the other end. Such \whs\ proved to be unstable
  \cite{stepan04, stepan05}, but their instability is of quite different
  nature than that described in this paper and in \cite{gon08}: it is
  related to a negative pole of the effective potential (\ref{Veff+}) at
  $u = 2k$ for Fisher's solution (\ref{ds+}), $\ep = +1$ and a similar
  singularity in its counterpart with an electric charge.

  These examples show that there remains quite a lot of work in studying the
  stability of more complicated solutions of scalar-tensor and $f(R)$
  theories.

\subsection*{Acknowledgements}

  We thank Roman Konoplya for helpful discussions. K.B. gratefully
  acknowledges the support of CAPES (Brazil) during his visit to UFES
  (Vit\'oria, ES, Brazil), where part of this work was done, and kind
  hospitality of the colleagues at UFES. The work of K.B. was also supported
  in part by RFBR grant 09-02-00677a, by NPK MU grant at PFUR, and by FTsP
  ``Nauchnye i nauchno-pedagogicheskie kadry innovatsionnoy Rossii'' for the
  years 2009-2013.
  A.Z. and J.C.F. were supported by Conselho Nacional de
  Desenvolvimento Cient\'ifico e Tecnol\'ogico (CNPq, Brazil).

\end{document}